\documentclass[journal]{IEEEtran}
\usepackage{amsmath,amsfonts}
\usepackage{algorithmic}
\usepackage{array}
\usepackage{subfigure}
\usepackage{textcomp}
\usepackage{stfloats}
\usepackage{url}
\usepackage{verbatim}
\usepackage{graphicx}
\usepackage{cite}
\usepackage{bm}
\begin{document}
	
	\title{Low-Complex Waveform, Modulation and Coding Designs for 3GPP Ambient IoT}
	
	\author{
		Mingxi Yin,
		Chao Wei,
		Kazuki Takeda,
		Yinhua Jia,
		Changlong Xu,
		Chengjin Zhang
		and Hao Xu,~\IEEEmembership{Fellow,~IEEE}
		\thanks{M. Yin, C. Wei, Y. Jia and C. Xu are with Qualcomm Wireless Communication Technologies (China) Limited, Beijing, 100013, China (e-mail: \{yinmx, weichao, yinhjia, changlon\}@qti.qualcomm.com).}
		\thanks{K. Takeda is with Qualcomm Japan GK, Tokyo, 107-0062, Japan (e-mail: ktakeda@qti.qualcomm.com).}
		\thanks{C. Zhang is with Qualcomm Technologies, Inc., San Diego, CA 92121-1714, USA (e-mail:chengjin@qti.qualcomm.com).}
		\thanks{H. Xu is with Qualcomm International, Inc., Beijing, 100013, China (e-mail: hxu@qti.qualcomm.com).}
		}

	\maketitle
	
	\begin{abstract}
		This paper presents a comprehensive study on low-complexity waveform, modulation and coding (WMC) designs for the 3rd Generation Partnership Project (3GPP) Ambient Internet of Things (A-IoT). A-IoT is a low-cost, low-power IoT system inspired by Ultra High Frequency (UHF) Radio Frequency Identification (RFID) and aims to leverage existing cellular network infrastructure for efficient RF tag management. The paper compares the physical layer (PHY) design challenges and requirements of RFID and A-IoT, particularly focusing on backscatter communications. An overview of the standardization for PHY designs in Release 19 A-IoT is provided, along with detailed schemes of the proposed low-complex WMC designs. The performance of device-to-reader link designs is validated through simulations, demonstrating 6 dB improvements of the proposed baseband waveform with coherent receivers compared to RFID line coding-based solutions with non-coherent receivers when channel coding is adopted.
	\end{abstract}
	
	
	\section{Introduction}	
	\IEEEPARstart{T}{he}
	Ambient Internet of Things (A-IoT) is a low-cost, low-power, and low-complexity IoT system developed under the 3rd Generation Partnership Project (3GPP) Study Item (SI)~\cite{ref_38848,ref_38769}. Inspired by Ultra High Frequency (UHF) Radio Frequency Identification (RFID)~\cite{ref_RFID_c1g2,ref_RFID_active} systems, 3GPP A-IoT aims to create an RF electronic tag system for applications such as warehousing, logistics, item tracking, and wireless sensor networks. Unlike traditional standalone and self-deployed RFID systems, 3GPP A-IoT leverages existing extensive access networks and terminals to communicate, manage and locate RF tags, referred to as A-IoT devices. This paper mainly focuses on the Release 19 SI for an overview on A-IoT waveform, modulation and coding (WMC) designs.
			
	Two key technologies from RFID have been incorporated into 3GPP A-IoT: energy harvesting~\cite{ref_EH} and backscatter communications~\cite{ref_backscatter,ref_coding}. 	
	Energy harvesting involves capturing ambient energy from various sources, such as solar, thermal, and RF signals, and converting it into usable electrical power. RF energy harvesting utilizes rectifiers, which convert RF signals into direct currents which further charge batteries or capacitors.	
	Backscatter modulation is a passive technique where the transmitter modulates backscatter coefficients on an external carrier wave (CW) to transmit information, without generating its own CW. The transmitter adjusts backscatter coefficients by switching among antenna load networks with different impedances~\cite{ref_Kurokawa}. This allows devices to transmit data without an internal RF source. The transmitted waveform, modulated by backscattering, is the product of backscatter coefficients and the external CW. 
	Besides RFID, backscatter modulations also have been studied in other systems, such as Wi-Fi~\cite{ref_wifi,ref_IQ}, Bluetooth~\cite{ref_ble} and LoRa~\cite{ref_lora}.	
			
	For the varying levels of complexity, capability and coverage, traditional RFID systems can be categorized into three main types: passive, semi-passive, and active.	
	Passive tags utilize both energy harvesting and backscatter communications. 
	Semi-passive tags are equipped with energy storage, e.g., batteries or capacitors, and communications still rely on backscatter.
	Active tags have energy storage and communicate actively with internal RF generation.	
	
\begin{figure}[!t]
		\centerline{\includegraphics[width=0.99\linewidth,scale=1.00]{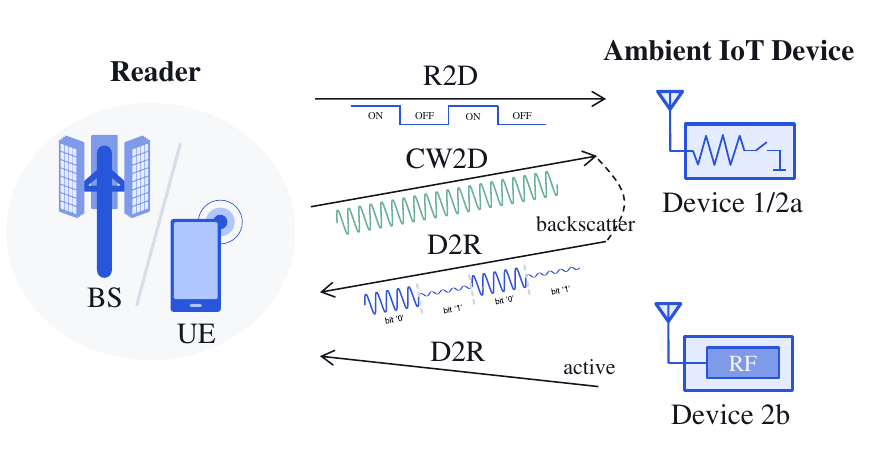}}
		\caption{System model for 3GPP A-IoT.}
		\label{fig_sys}
	\end{figure}

	Similarly, A-IoT devices have three primary categories in Release 19 SI: Device 1, Device 2a, and Device 2b. The overall system model for 3GPP A-IoT is illustrated in Fig.~\ref{fig_sys}. All device types are assumed to have energy storage with energy harvesting capabilities. Device 1 consumes around 1 $\mu$W, while Devices 2a and 2b consume hundreds of $\mu$W. 
	Device 1 and Device 2a are designed for backscatter communications, with Device 1 being ultra-low-power and Device 2a having higher power consumption with an amplifier. Device 2b, however, uses non-backscatter methods, requiring more power but offering higher performance. A unified physical layer (PHY) solution is necessary in A-IoT to ensure compatibility across all device types, aligning the system to the lower capability type with backscatter communications. 
	
	The system model of A-IoT is illustrated in Fig.~\ref{fig_sys}. A base station (BS) or user equipment (UE) can acts as a reader for A-IoT devices.
	A-IoT devices receive commands from readers via the reader-to-device (R2D) channel. For Device 1 and 2a, which use backscatter communications, the transmitting channel includes two segments: the carrier wave-to-device (CW2D) and the device-to-reader (D2R) channels. In contrast, Device 2b, which is non-backscatter, does not have CW2D channel. In A-IoT, PHY channels of R2D and D2R links are referred to as PRDCH and PDRCH, respectively.
	The backscatter communication system illustrated in Fig.~\ref{fig_sys} operates in a monostatic way, where the CW emitter and the reader are co-located, as also commonly used in RFID systems. Additionally, bistatic backscatter is also supported in A-IoT, where the CW emitter and the reader are separate.
	
	Traditional UHF RFID systems, especially passive ones, primarily operate in line-of-sight (LOS) environments. When LOS paths are obstructed, RFID tags struggle to connect with readers. Additionally, the limited receiver sensitivity of tags and the simplistic physical layer design restrict the coverage, typically to less than 10 meters, making passive RFID suitable mainly for indoor scenarios. 	
	A-IoT targets indoor environments with a larger coverage of up to 30 meters, considering fast fading channels with and without LOS. Therefore, A-IoT needs to enhance physical layer designs for the improved coverage target.
	
	This paper addresses low-complex WMC designs for A-IoT. The contributions of this paper are threefold:
	\begin{itemize}
		\item[$\bullet$] Compare the different PHY challenges and requirements of RFID and 3GPP A-IoT. Summarize the waveform, modulation and coding studies of the recent A-IoT. 
		\item[$\bullet$] Further illustrate details of the proposed low-complexity waveform, modulation and coding designs for A-IoT.
		\item[$\bullet$] Verified with link-level simulations for D2R, the proposed square-wave baseband modulations show a 3-6 dB error performance gain compared to RFID line coding solutions when channel coding is adopted.
	\end{itemize}

	\section{RFID versus A-IoT in PHY}	
	PHY designs of RFID, which is a well-established commercial system, can provide valuable insights for A-IoT. RFID systems utilize various waveforms, modulation techniques and coding schemes depending on link directions. In this context, the passive UHF RFID standard EPC-C1G2~\cite{ref_RFID_c1g2} is used as the primary reference to represent RFID systems.	
	RFID systems currently face several limitations, including low reliability, short coverage, and a strong dependence on LOS. 
	This section explores how A-IoT can build upon the foundational principles of RFID while addressing its shortcomings to achieve more robust and versatile applications. 	

	\subsection{Overview on the Existing RFID WMC}		
	The block diagram for PHY of UHF RFID is illustrated in Fig.~\ref{fig_flow}(a), depicting two link directions: R$\Rightarrow$T and T$\Rightarrow$R, which correspond to A-IoT R2D and D2R, respectively. RFID exhibits a tighter link budget in the R$\Rightarrow$T compared to the T$\Rightarrow$R if simply considering free-space propagation. This is due to the poor sensitivity of tag receivers, e.g., $-20$ dBm, which is significantly lower than the sensitivity of reader receiving T$\Rightarrow$R signals, e.g., $-90$ dBm. Overall, given the receiver sensitivities, noise figures and characteristics of the backscatter communication, both R$\Rightarrow$T and T$\Rightarrow$R of passive RFID operate in the high signal-to-noise ratio (SNR) regime. Consequently, WMC designs for RFID do not pursuit high error performance.
	
	\subsubsection{Waveform and Modulation}	
	In RFID R$\Rightarrow$T link, Amplitude Shift Keying (ASK), e.g., On-Off Keying (OOK), or Phase-Reversal ASK (PR-ASK) is adopted for modulation and can be demodulated by tags with envelope detectors. 	
	In RFID T$\Rightarrow$R link, ASK or Phase Shift Keying (PSK) backscatter modulation is used. ASK backscatter modulation maps information bits through different amplitudes of backscatter coefficients, whereas PSK backscatter modulation utilizes phases.	
	The primary distinction between ASK and PSK backscatter modulation lies in their energy efficiencies, with PSK backscatter modulation achieving higher energy efficiency~\cite{ref_backASKPASK}. However, impedance matching for PSK is more challenging, requiring more complex hardware implementations.
	
	\subsubsection{Line Coding}	
		
	RFID utilizes different line coding schemes for R$\Rightarrow$T and T$\Rightarrow$R. In some aspects, line coding is more closely related to modulation concept than to modern coding, as it is neither source coding that compresses information nor channel coding that checks or corrects errors. Instead, it generates self-synchronizing waveforms for low-complexity systems by embedding clock information within each bit with symbol transitions. There are multiple line codes involved in RFID as illustrated in Fig.~\ref{fig_wave}(a).	
	
	RFID R$\Rightarrow$T uses Pulse Interval Encoding (PIE), which maps bits 0 and 1 to unequal-length OOK waveforms. Another active RFID system~\cite{ref_RFID_active} uses Manchester coding for R2D, which maps bits 0 and 1 to opposite ``high''-``low'' level transition directions in the middle. PIE has a slight energy efficiency advantage over Manchester coding, making it more suitable for passive tags. However, Manchester coding offers the better communication performance. 
	
	Passive RFID employs two types of line codes for T$\Rightarrow$R: FM0 and Miller-modulated subcarrier (MMS). Both FM0 and Miller are state-transition-based coding, similar to convolutional coding (CC). However, due to their simple state transition rules, they lack error-correcting capabilities. FM0, also known as bi-phase space encoding, can be understood as a frequency modulation where the waveform for bit 1 is flat, and the waveform for bit 0 has a frequency by involving a level transition. A similar code is FM1, which frequency modulates only bit 1. MMS involves multiplying the Miller-encoded signal with a square wave (i.e, referred to as a subcarrier in~\cite{ref_RFID_c1g2}), which contributes to a frequency shift away from the incident CW. The frequency shift separates the CW and D2R signals in the spectrum, preventing CW interference. Additionally, waveforms of both FM0 and MMS have inter-bit waveform correlations, i.e., the differential characteristic. The ``high'' and ``low'' levels in line codes are then mapped to ASK or PSK backscatter modulation symbols as shown in Fig.~\ref{fig_wave}(c).	
	
	The reception of FM0 and MMS signals can be achieved using either coherent or non-coherent receivers. Although different literature works may slightly vary in their definitions of coherent and non-coherent receivers, they can generally be distinguished as follows:	
	\begin{itemize}
		\item[$\bullet$] Coherent Receiver: This type of receiver first performs channel estimation and then demodulates using absolute modulation symbols. FM0 and MMS signals can be decoded using the Viterbi decoder with soft decisions for optimal performance.
		\item[$\bullet$] Non-Coherent Receiver: This receiver does not require channel estimation. It detects the transition rules in received waveforms rather than knowing absolute symbols. FM0 and MMS signals can be decoded by symbol transition detection in the middle or correlation detection using known patterns. 
	\end{itemize}		
	In summary, coherent receivers can achieve better error performance but are more complex. In scenarios where the T$\Rightarrow$R link budget is sufficient, RFID readers can choose non-coherent receivers to reduce complexity and costs.	
	
	\begin{figure}[!t]
		\centerline{\includegraphics[width=0.99\linewidth,scale=1.00]{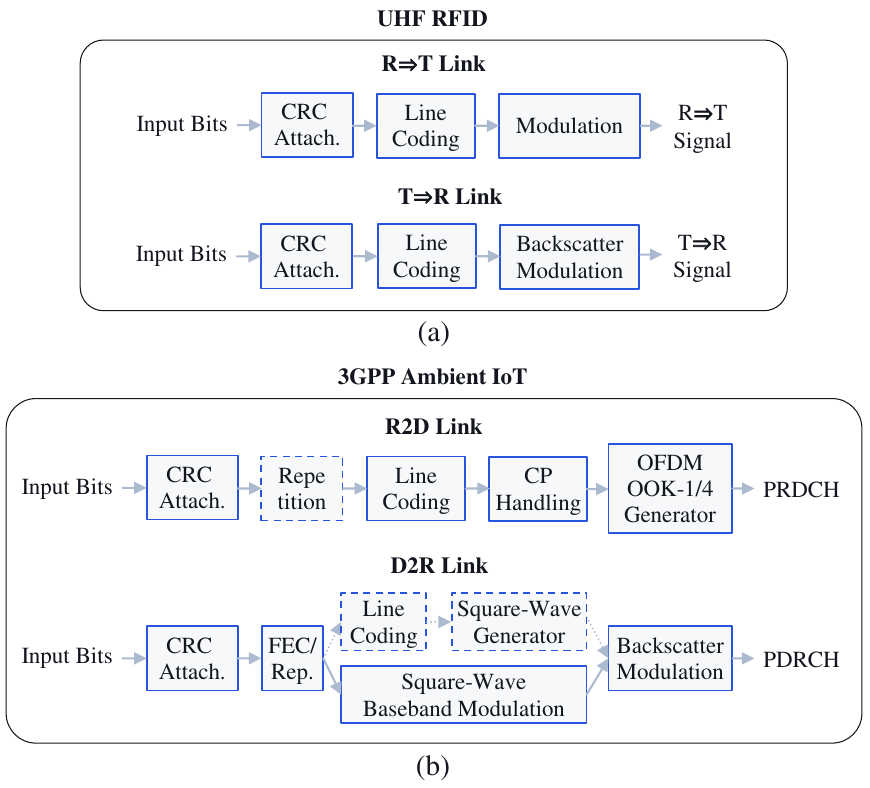}}
		\caption{Block diagram for PHY of UHF RFID and 3GPP A-IoT.}
		\label{fig_flow}
	\end{figure}	
	
	\subsubsection{Cyclic Redundancy Check (CRC)}	
	CRC-5 and CRC-16 are used for different signaling types in R$\Rightarrow$T and T$\Rightarrow$R.
	
	Other PHY techniques such as interleaving and scrambling are not adopted in RFID due to the complexity limitation.		
	
	\subsection{WMC Enhancements in A-IoT}	
	In this context, significant differences between A-IoT and RFID arise in terms of D2R communications due to the improved link budget target. To extend coverage, A-IoT firstly assumes a 10 dB improvement in receiver sensitivity for R2D compared to passive RFID. Despite this sensitivity improvement, R2D still operates in the high SNR regime. However, this results in a 10 dB budget loss for backscatter-based D2R. Additionally, given that the receiving sensitivities of BS/UE are higher than that of RFID readers, A-IoT D2R could operate in the middle or low SNR regime. This shift directly increases the importance of D2R waveform, modulation and coding schemes in A-IoT.	

	For R2D, a primary challenge in A-IoT is ensuring waveform compatibility with the existing NR waveform orthogonal frequency division multiplexing (OFDM) waveforms, which have their own frame structure, such as a cyclic prefix (CP). A-IoT readers, which are either BSs or UEs, are already equipped with OFDM waveform generators which can be reused for A-IoT R2D. However, the challenges associated with CP structure should be carefully handled in A-IoT, such as the creation of false edges and the impact on decoding of PIE or Manchester. 

	For D2R, a key issue is enhancing link performance to achieve the larger coverage target. One approach is targeting coherent receivers to achieve optimal performance. This directly motivates the inclusion of channel coding, such as Forward Error Correction (FEC) coding, in A-IoT D2R. Furthermore, regarding waveform and modulation, given the use of coherent receivers, a primary question is whether we should replicate the passive RFID approach. The answer likely no, because FM0 or MMS line coding in RFID are not optimal-performance solutions designed for coherent receivers. Due to the involvement of FEC, the joint performance of waveform and modulation concatenated with FEC should also be considered.	
	An overall diagram for A-IoT PHY processing is shown in Fig.~\ref{fig_flow}(b), where detailed WMC designs are given in the following sections.	
	
	\section{Waveform Generation}
	This section summarizes the 3GPP A-IoT designs for waveform and modulation in R2D and D2R separately, with detailed highlights of our WMC proposals.
	
	\subsection{R2D}	
	A-IoT R2D adopts the OOK modulation, similar to RFID. 
	Utilizing existing OFDM transmitters offers the benefit of cost reduction for R2D waveform generation. To this end, the 3GPP A-IoT study has drawn insights from another 3GPP topic, the low-power wake-up signal (LP-WUS)~\cite{ref_LPWUS}, which addresses the compatibility of OFDM waveforms with OOK. Specifically, there are two types of waveforms referred to as OOK-1 and OOK-4, differing primarily in their data rates.	
	OOK-1 carries one OOK symbol per OFDM symbol, representing the OOK ON and OFF states through the high and low energy levels of the entire OFDM symbol. In contrast, OOK-4 can carry multiple OOK symbols per OFDM symbol with a higher symbol rate, with calculating frequency domain signals for OFDM based on the desired OOK time-domain waveform, such as DFT-s-OFDM or least square (LS) algorithm.  
		
	Regarding line codes for R2D waveform, 3GPP A-IoT Release 19 primarily discussed two schemes in the R2D link: PIE and Manchester. Manchester has garnered more support, mainly due to its superior communication performance and the regularity of its equal-length 0 and 1 bits, which better align with the NR frame structure and resource scheduling. 
	
	Additionally, using existing OFDM transmitters requires addressing the issue of CP disruption to the Manchester (or PIE) coded OOK waveform, such as the creation of false edges which do not satisfy Manchester rule.
	We proposed a low-complex solution that retains subcarrier orthogonality by copying the starting OOK chip(s) and inserting their inversions at the end of the OFDM symbol as check chips before CP insertion. Consequently, the CP is identical to and further merged with the starting chip. Additionally, the starting OOK chip can be shortened to ensure that the OOK chips after CP insertion are of nearly equal length. Hence, the final waveform still follows Manchester rule.
	
	\subsection{D2R}	
\begin{figure}[!t]
		\centerline{\includegraphics[width=0.99\linewidth,scale=1.00]{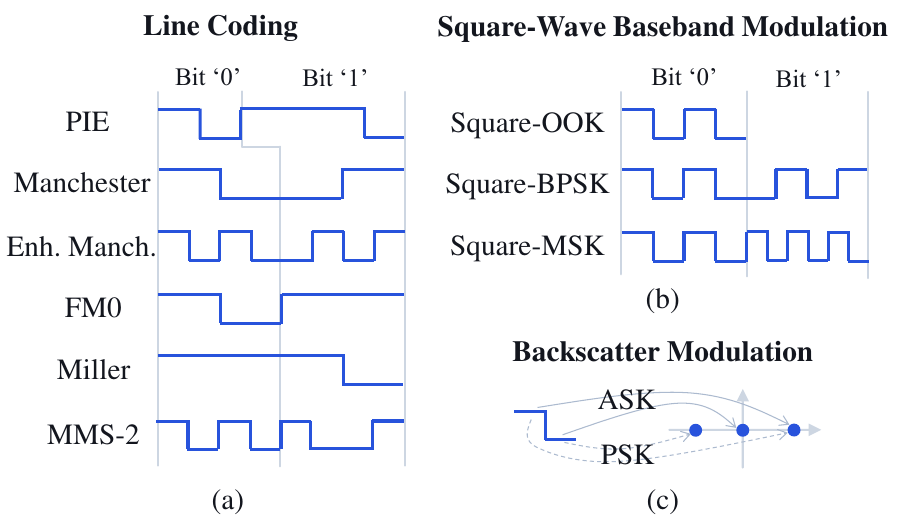}}
		\caption{Waveforms in RFID and A-IoT. (a) Line codes in RFID and A-IoT studies; (b) Square-wave modulation based baseband waveforms for A-IoT; (c) Illustration for ASK/PSK backscatter modulation.}
		\label{fig_wave}
	\end{figure}

	In Release 19 SI, there are three modulation options discussed for A-IoT D2R: ASK, PSK and frequency shift keying (FSK).
	Where ASK and PSK are backscatter modulations given in Fig.~\ref{fig_wave}(c). However, FSK requires baseband modulation under laying backscatter modulations. Minimum Shift Keying (MSK), a kind of FSK, is shown in Fig.~\ref{fig_wave}(b) labeled by ``square-MSK''. The frequency shift is implemented by switching the frequencies of backscatter coefficients, which requires a baseband waveform. Thus, the baseband waveform modulation based on square waves is proposed under-laying backscatter modulations, as shown at the bottom row of Fig.~\ref{fig_flow}(b). Similarly, ASK and PSK modulations on the baseband square waveform can be also supported, such as the ``square-OOK'' and ``square-BPSK'' examples shown in Fig.~\ref{fig_wave}(b).	
	
	The baseband square-wave ASK modulation (``square-OOK''), can be simply implemented by presence and absence of square waves. 
	Baseband square-wave PSK modulation carries information through the initial phase of the square wave corresponding to each bit group. For example, in square wave BPSK modulation, a bit 0 corresponds to a square wave with the initial phase 0, while a bit 1 corresponds to the initial phase $\pi$. In square wave QPSK modulation, the 2-bit 00, 01, 10, and 11 correspond to initial phases of 0, $\pi/2$, $\pi$, and $3\pi/2$, respectively.
	Baseband square-wave FSK modulation, such as binary-FSK (BFSK), transmits two different frequencies of square waves for bits 0 and 1. MSK is considered for its high spectrum efficiency among BFSK techniques. It is implemented using two frequencies of square waves with a 1.5x relationship representing bits 0 and 1, respectively, with phase continuity in square waves across bits.	
	Baseband waveforms are finally generated with ``high'' and ``low'' levels of square waves mapped to ASK or PSK backscatter modulation symbols as shown in Fig.~\ref{fig_wave}(c).
	The square-wave ASK/PSK/FSK baseband modulation schemes can achieve the same error performance of classic ASK/PSK/FSK modulations with coherent receivers.
	
	One benefit of the proposed square-wave baseband modulation is that the waveform incorporates square waves per bit, which include backscatter modulation symbol inversions similar to those in line coding, thereby providing clock information.		
	Another motivation of square waveform is making D2R signals frequency shifted from CW using square-wave frequencies to avoid CW interference. 
	
	Additionally, as a parallel solution of the proposed baseband square-wave modulation, A-IoT also considered frequency-shift capable line coding, such as the enhanced Manchester and MMS, as shown at the bottom of Fig.~\ref{fig_flow}(b). The enhanced Manchester waveform that illustrated in Fig.~\ref{fig_wave}(a), is traditional Manchester plus a square-wave generator that also used in MMS, to support frequency shifts. Interestingly, the enhanced Manchester and the proposed square wave BPSK modulation can produce the same waveform, as Manchester can also be viewed as phase modulation of a baseband square wave. However, square wave modulation is more flexible in supporting various modulation schemes such as baseband ASK, PSK, and MSK modulation. 
	In Release 19 SI discussions, line coding and MSK modulation were considered as two separate steps. However, by examining their waveforms shown in Fig.~\ref{fig_wave}, the redundancy in this concatenation becomes evident. Furthermore, the necessity of involving baseband modulation parallel to line coding is demonstrated.
	
\begin{figure}[!t]
		\centerline{\includegraphics[width=0.98\linewidth,scale=1.00]{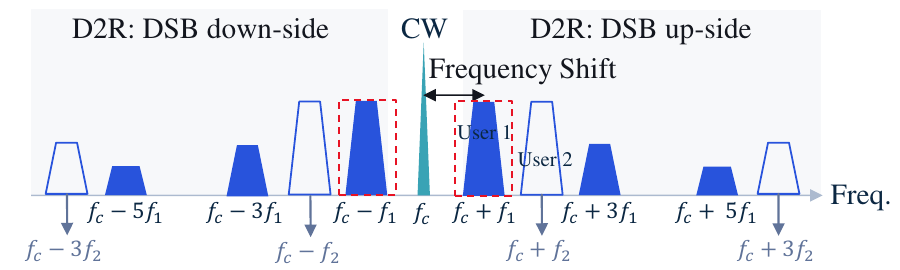}}
		\caption{Illustration for D2R waveform spectrum to allow FDMA.}
		\label{fig_D2Rspectrum}
	\end{figure}

	The spectrum for the square-wave baseband modulation is illustrated in Fig.~\ref{fig_D2Rspectrum}. The frequency of square wave determines the distance by which the D2R signal center is shifted from the carrier spectrum. 
	The shown spectrum is double-sideband (DSB), with mirrored spectrum on both sides of CW. Single-sideband (SSB) is also possible for Device 1/2a with increased hardware complexity~\cite{ref_IQ}, while can be more easily supported by Device 2b.
	
	For the receiver, modulated square waves can be viewed as modulated sinusoidal waves (containing the majority proportion of the energy) plus a set of higher-order harmonic components. If a larger number of backscatter modulation order is supported, the square waves can be further refined to multi-stage waves~\cite{ref_lora} to suppress harmonics. 
	The receiver can filter out the sinusoidal frequency part shown as the red dash boxes in Fig.~\ref{fig_D2Rspectrum}. Additionally, the receiver can utilize a larger bandwidth to combine the higher-order harmonics for demodulation, thereby improving the performance. 
	Device 2b can directly generate the modulated sinusoidal waveform without high-order harmonics, which is compatible with square-wave modulations. So that this design is unified to different device types. 

	In addition, the square wave baseband waveform facilitate Frequency Division Multiple Access (FDMA) by utilizing their frequency shifting properties. We leverage the fact that square wave harmonics only appear at odd harmonics (1st, 3rd, 5th, etc.), allowing us to place the signal of another device at the even harmonic positions of the first device. An FDMA example with two users is illustrated in Fig.~\ref{fig_D2Rspectrum}, where User 2's square wave frequency $f_2$ is twice that of User 1 $f_1$. Iteratively, the 3rd, 4th, and subsequent users can adopt $4, 8, \cdots$ multiples of $f_1$. In general, D2R FDMA operates by allowing other users to use even multiples of the first user's square-wave frequency.
	However, square wave-enabled FDMA faces challenges due to the frequency accuracy of square waves generated by A-IoT devices. The low clock accuracy of A-IoT Device 1/2a causes frequency shifts with uncertainties, which could increase interference among FDMA users.

	\section{Coding Designs}	
	\subsection{FEC Coding}
	In R2D, no FEC coding but only repetition is considered at this stage. The primary reason for not considering FEC is the high decoding complexity for tags. Additionally, envelope detection receivers of tags cannot effectively utilize FEC performance due to their inability to support soft decisions. 
	
	In D2R, A-IoT has introduced significant changes by incorporating convolutional codes (CC) as FEC compared to RFID. The encoding complexity of FEC is relatively low, especially for CC, whose encoders are based on shift register designs. This complexity is similar to the CRC encoding complexity that already supported by passive RFID. 
	
	CC has been employed in 3GPP standards since the 3rd generation (3G). CC with a rate of $R = \frac{1}{2}$ and a constraint length $K = 9$ was utilized in 3G. Then in Long Term Evolution (LTE), CC $[133, 171, 165]$ was adopted with $R = \frac{1}{3}$ and $K = 7$. The shorter constraint length results in an exponential reduction in decoding complexity. Thus, the lower code rate is adopted to compensate for the performance loss. 
	
	To support varying code rates, nested CC polynomials are recommended. Herein, two design directions can be considered and the nested CC polynomial groups are searched with good error performance.
	\begin{itemize}
		\item[$\bullet$] Maintain a constraint length of $K = 7$ and develop nested polynomials based on the LTE's CC for lower code rates. 
		\begin{itemize}
			\item[-] Nested CC polynomials for code rates $1/2$ to $1/4$: a. $[133, 171, 165, 117]$ (adopted in~\cite{ref_wimax}), b. $[133, 171, 165, 127]$ or c. $[133, 171, 165, 137]$.
		\end{itemize}
		\item[$\bullet$] Reduce the constraint length to $K = 6$ and study new nested polynomials of varying code rates. This reduction could further decrease decoding complexity, while considering further lower code rates to ensure coding gains.
		\begin{itemize}
			\item[-] Nested CC polynomials for code rates $1/2$ to $1/6$: a. $[45, 73, 75, 67, 57, 55]$, b. $[55, 67, 77, 51, 53, 73]$ or c. $[77, 73, 55, 45, 67, 65]$.
		\end{itemize}
	\end{itemize}
		
	\textbf{\textit{Complexity Analysis}}: The encoding complexity for CC with a constraint length of 6 is slightly lower than that for CC with a constraint length of 7, as the former requires 5 shift registers, while the latter requires 6 shift registers. 
	However, the decoding complexity for CC with a constraint length of 6 is half that of CC with a constraint length of 7, given the same code rate.
	
	Conventional interleaving requires memory to store coded bits temporarily, which is too demanding for A-IoT devices, especially Device 1. Thanks to the polynomial-based nature of CC, a memory-free interleaving can be implemented using polynomial-sweeping based CC encoding, eliminating the need for an additional buffer for coded bits and simplifying the design. Specifically, for a CC with $n$ polynomials, this memory-free interleaving is achieved by performing $n$ encoding operations, with only the $i$-th polynomial active during the $i$-th encoding.
	
	\subsection{Nested CRC Design}
	In CRC design, two main factors need to be balanced: protection performance (such as undetected errors and false alarms), and overhead. The performance of CRC checks is exponentially related to the CRC length $L$ (approximately $2^{(L-1)}$), while the longer CRC length increases overhead.
	Therefore, multiple CRC lengths for different data lengths can be considered. A nested CRC structure is recommended to reduce the hardware complexity, by allowing different length CRC to share a single CRC generator to minimize the number of polynomials.
	
	Three CRC lengths 6, 11, and 16 can be considered in A-IoT to balance performance and overhead. Two nested CRC designs can be considered. One approach is to design nested CRC-11 and CRC-16 based on the existing CRC-6 used in NR and the nested CRC polynomial is $x^{16}+x^{11}+x^6+x^5+1$. The other approach is to search for a new set of nested CRC with better performance, such as $x^{16}+x^{11}+x^6+x^4+x^3+1$.
	For example, given 128 input bits, CRC-16 of the NR-based nested CRC achieves an undetected error probability of $2.4 \times 10^{-6}$, while CRC-16 of the new search nested CRC achieves $1.4 \times 10^{-6}$, as evaluated in simulations.
	
	\textbf{\textit{Complexity Analysis}}: If adopt two CRC lengths with non-nested structures, e.g., reusing NR CRC-6 and CRC-16, it requires a total of 24 shift registers and 5 XOR operators. However, the proposed nested CRC polynomial for three lengths 6, 11 and 16 only requires one CRC generator with only 16 shift registers and 4 to 5 XOR operators.
	
	\section{Performance Evaluation}	
	This section evaluates the link-level performance of A-IoT with simulations. Due to the high similarity with RFID in the R2D link, the focus is on the WMC and FDMA performance of A-IoT D2R, which has been enhanced compared to RFID. 	
	
	\subsection{Evaluation for D2R WMC}	
\begin{figure}[!t]
		\centering
		\subfigure[Evaluation on the nested CC]{
			\begin{minipage}[!t]{0.463\linewidth}
				\flushleft
				\includegraphics[width=1\linewidth,scale=1.00]{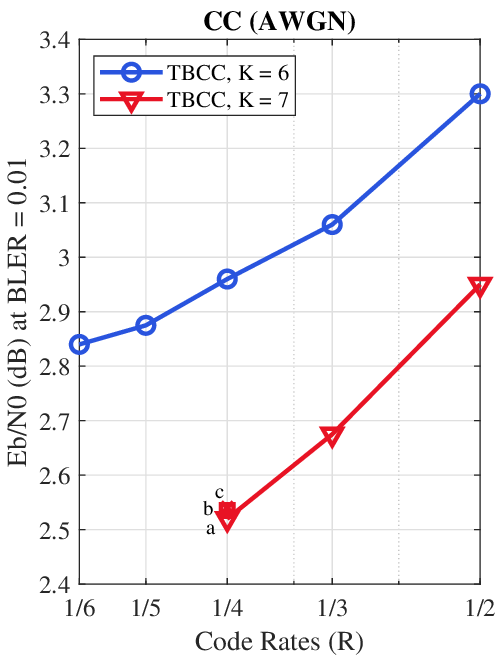}
			\end{minipage}
			\label{fig_CC_BLER}
		}%
		\subfigure[Evaluation on PDRCH]{
			\begin{minipage}[!t]{0.49\linewidth}
				\flushleft
				\includegraphics[width=1\linewidth,scale=1.00]{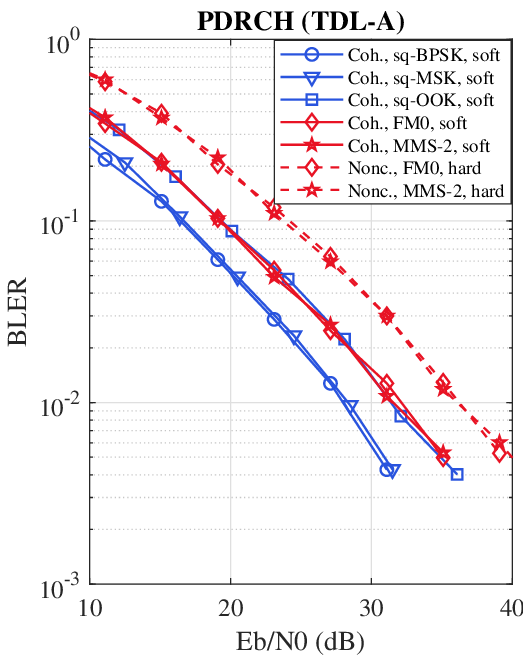}
			\end{minipage}
			\label{fig_D2R_BLER}
		}%
		\caption{BLER performance for A-IoT D2R, with the information block size of 128 bits: (a) BLER for the nested CC with $K=6,7$ using tail-biting CC (TBCC) with AWGN channel and BPSK (soft decisions); (b) BLER for overall PDRCH over TDL-A channel (30 ns delay spread, 3 km/h) with various waveform schemes. The CC $[133,171]$ is used in all cases. CW is single-tone at 900 MHz, frequency shift is 240 kHz for square-wave BPSK/MSK/OOK modulations, bit rate is 60 kbps, backscatter modulation is BPSK. CW2D and D2R are independent TDL-A channels. The reader is 1Tx-1Rx, with coherent or non-coherent receivers using soft or hard decisions for CC decoding, respectively. Perfect channel knowledge is assumed.}
		\label{fig_WMC}
	\end{figure}

	The Fig.~\ref{fig_WMC} captures the block error rate (BLER) performance for the A-IoT D2R WMC, with the FEC and overall PDRCH performance plotted in (a) and (b), separately.
	
	Coding gains of the nested CC with constraint lengths $K=6$ and $7$ are compared in Fig.~\ref{fig_CC_BLER}. For $K=7$, the three polynomial options (``a'', ``b'', and ``c'') for code rates $R = 1/4$ exhibit nearly identical BLER performance. For $K = 6$, the figure only presents the varying rates of the nested polynomial set ``a'' because the three options demonstrate comparable performance. Overall, the nested CC with $K = 7$ outperforms $K = 6$ by coding gains of 0.3-0.4 dB at a BLER of 1\% under the same code rates, while $K = 6$ reduces decoding complexity by half.
	
	The overall performance for PDRCH including D2R waveform, modulation and coding, using coherent and non-coherent receivers, are plotted in Fig.~\ref{fig_D2R_BLER}. Two parallel solutions given in Fig.~\ref{fig_flow} are compared, i.e., CC concatenated with either the proposed square-wave baseband modulations or line codes. When CC is used as FEC in all curves, the BLER performance with square-wave baseband modulations and a coherent receiver significantly outperforms those using FM0 and MMS-2 line codes with either a coherent or non-coherent (correlation detection) receiver by 3 dB and 6 dB, respectively.
	This is because the correlation between waveforms of consecutive bits in the FM0 and MMS line codes degrades the error correction performance of CC. Notably, the enhanced Manchester code performs identically to square-BPSK due to their identical waveforms.	
	In summary, under the assumption of using CC and a coherent receiver, line coding appears redundant and ineffective. Square-wave baseband modulations offer better error performance and greater flexibility.
	
	\subsection{Evaluation for D2R FDMA}
\begin{figure}[!t]
		\centering
		\includegraphics[width=0.86\linewidth,scale=1.00]{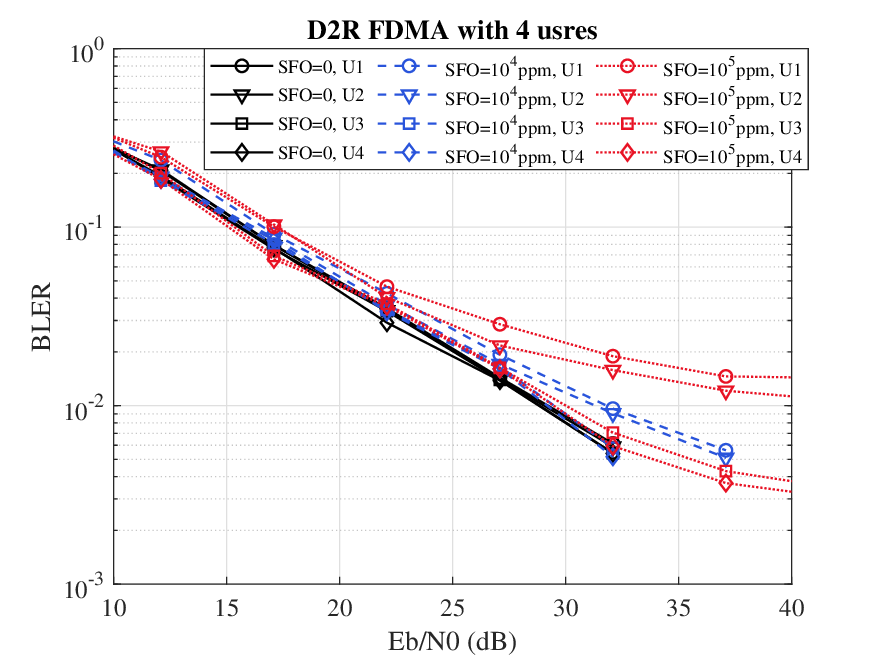}
		\caption{BLER performance for A-IoT D2R FDMA with 4 users enabled by square-wave baseband waveforms over TDL-A channel (30 ns delay spread, 3 km/h). The square wave frequencies of 4 users are 60 Hz, 120 Hz, 240 Hz and 480 Hz, respectively. All users use the same bit rate of 7.5 kbps, CC $[133,171]$, square-wave BPSK modulations and BPSK backscatter modulation. The SFO = $10^4$ ppm is assumed for Device 1 and $10^5$ ppm for Device 2a. Other assumptions are same to Fig.~\ref{fig_D2R_BLER}.}
		\label{fig_FDMA_BLER}
	\end{figure}

	The BLER performance of A-IoT D2R FDMA are plotted in Fig.~\ref{fig_FDMA_BLER}, where four A-IoT devices concurrently transmit D2R signals using the even-multiple based baseband square-wave frequencies. These devices are assumed to experience sampling clock frequency offsets (SFOs) due to low clock accuracy, leading to uncertain frequency shifts and subsequent interference among FDMA users. 
	The results demonstrate that a smaller SFO of $10^4$ ppm has a negligible impact, closely matching the performance observed with an ideal clock (SFO = 0 in Fig.~\ref{fig_FDMA_BLER}). In contrast, an SFO of $10^5$ ppm degrades error performance and even results in error floors. Consequently, a low-complexity device clock calibration scheme that maintains residual SFO around $10^4$ ppm for D2R transmission is highly desirable for D2R FDMA.

	\section{Conclusions}
	This work addresses the critical need for low-complexity PHY designs in 3GPP A-IoT systems. Similarities and differences between RFID and A-IoT are compared, WMC designs in 3GPP A-IoT Release 19 SI are summarized, and the proposed novel WMC schemes for A-IoT are highlighted. In R2D, CP insertion is handled by check chips to reuse OFDM transmitters. In D2R, the square-wave baseband modulations and the nested CC schemes provide improved performance with coherent receivers. FDMA in D2R is supported by square-wave waveforms with even multiples of frequency. The nested CRC design further reduces overhead with low complexity. 	When CC is adopted to improve D2R link error performance, the simulation results demonstrate a 6 dB gain of the proposed square-wave baseband modulation with the coherent receiver compared to RFID line coding with the non-coherent receiver.

\end{document}